\documentclass[amsmath,amssymb,onecolumn,pra,superscriptaddress]{revtex4}
\usepackage{amsfonts}
\usepackage{amsmath}
\usepackage{amsthm}
\usepackage{amscd}
\usepackage{amssymb}
\usepackage{subfigure}
\usepackage{amsxtra}
\usepackage{gensymb}
\usepackage{bm}           
\usepackage{bbm}
\usepackage{graphicx}
\usepackage{epstopdf}
\usepackage{color}
\usepackage{times}
\usepackage{mdframed}
\usepackage[colorlinks=true,citecolor=blue,urlcolor=black]{hyperref}

\def\bi#1\ei {\begin{itemize}#1\end{itemize}}
\def\bn#1\en {\begin{enumerate}#1\end{enumerate}}
\def\bea#1\eea {\begin{align}#1\end{align}}
\def\bean#1\eean {\begin{align*}#1\end{align*}}
\def\ben#1\een {\begin{equation*}#1\end{equation*}}
\def\be#1\ee {\begin{equation}#1\end{equation}}
\def\bes#1\ees {\begin{equation}\begin{split}#1\end{split}\end{equation}}
\def\bear#1\eear {\begin{eqnarray}#1\end{eqnarray}}
\def\bear#1\eear {\begin{eqnarray*}#1\end{eqnarray*}}

\emergencystretch=\maxdimen
\newcommand{\beq}{\begin{equation}}
\newcommand{\eeq}{\end{equation}}

\newcommand{\ket}[1]{\ensuremath{\left|#1\right\rangle}}

\newcommand{\blue}{\color{blue}}

\begin{document}

\title{\bf Experimental measurement-device-independent quantum digital signatures over a metropolitan network}

\author{Hua-Lei Yin}
\affiliation{Hefei National Laboratory for Physical Sciences at Microscale and Department of Modern Physics, University of Science and Technology of China, Hefei, Anhui 230026, China}
\affiliation{CAS Center for Excellence and Synergetic Innovation Center in Quantum Information and Quantum Physics, University of Science and Technology of China, Hefei, Anhui 230026, China}

\author{Wei-Long Wang}
\affiliation{EI Telecomunicaci$\acute{o}$n, Department of Signal Theory and Communications, University of Vigo, Vigo E-36310, Spain}

\author{Yan-Lin Tang}
\affiliation{Hefei National Laboratory for Physical Sciences at Microscale and Department of Modern Physics, University of Science and Technology of China, Hefei, Anhui 230026, China}
\affiliation{CAS Center for Excellence and Synergetic Innovation Center in Quantum Information and Quantum Physics, University of Science and Technology of China, Hefei, Anhui 230026, China}

\author{Qi Zhao}
\affiliation{Center for Quantum Information, Institute for Interdisciplinary Information Sciences, Tsinghua University, Beijing, 100084, China}

\author{Hui Liu}
\author{Xiang-Xiang Sun}
\affiliation{Hefei National Laboratory for Physical Sciences at Microscale and Department of Modern Physics, University of Science and Technology of China, Hefei, Anhui 230026, China}
\affiliation{CAS Center for Excellence and Synergetic Innovation Center in Quantum Information and Quantum Physics, University of Science and Technology of China, Hefei, Anhui 230026, China}

\author{Wei-Jun Zhang}
\author{Hao Li}
\affiliation{State Key Laboratory of Functional Materials for Informatics, Shanghai Institute of Microsystem and Information Technology, Chinese Academy of Sciences, Shanghai 200050, China}

\author{Ittoop Vergheese Puthoor}
\affiliation{SUPA, Institute of Photonics and Quantum Sciences, Heriot-Watt University, Edinburgh EH14 4AS, United Kingdom}

\author{Li-Xing You}
\affiliation{State Key Laboratory of Functional Materials for Informatics, Shanghai Institute of Microsystem and Information Technology, Chinese Academy of Sciences, Shanghai 200050, China}

\author{Erika Andersson}
\affiliation{SUPA, Institute of Photonics and Quantum Sciences, Heriot-Watt University, Edinburgh EH14 4AS, United Kingdom}

\author{Zhen Wang}
\affiliation{State Key Laboratory of Functional Materials for Informatics, Shanghai Institute of Microsystem and Information Technology, Chinese Academy of Sciences, Shanghai 200050, China}

\author{Yang Liu}
\author{Xiao Jiang}
\affiliation{Hefei National Laboratory for Physical Sciences at Microscale and Department of Modern Physics, University of Science and Technology of China, Hefei, Anhui 230026, China}
\affiliation{CAS Center for Excellence and Synergetic Innovation Center in Quantum Information and Quantum Physics, University of Science and Technology of China, Hefei, Anhui 230026, China}

\author{Xiongfeng Ma}
\affiliation{Center for Quantum Information, Institute for Interdisciplinary Information Sciences, Tsinghua University, Beijing, 100084, China}
\affiliation{CAS Center for Excellence and Synergetic Innovation Center in Quantum Information and Quantum Physics, University of Science and Technology of China, Hefei, Anhui 230026, China}

\author{Qiang Zhang}
\affiliation{Hefei National Laboratory for Physical Sciences at Microscale and Department of Modern Physics, University of Science and Technology of China, Hefei, Anhui 230026, China}
\affiliation{CAS Center for Excellence and Synergetic Innovation Center in Quantum Information and Quantum Physics, University of Science and Technology of China, Hefei, Anhui 230026, China}

\author{Marcos Curty}
\affiliation{EI Telecomunicaci$\acute{o}$n, Department of Signal Theory and Communications, University of Vigo, Vigo E-36310, Spain}

\author{Teng-Yun Chen}

\author{Jian-Wei Pan}

\affiliation{Hefei National Laboratory for Physical Sciences at Microscale and Department of Modern Physics, University of Science and Technology of China, Hefei, Anhui 230026, China}
\affiliation{CAS Center for Excellence and Synergetic Innovation Center in Quantum Information and Quantum Physics, University of Science and Technology of China, Hefei, Anhui 230026, China}

\date{\today}

\begin{abstract}
Quantum digital signatures (QDS)
provide a means for signing electronic communications with information-theoretic security. However, all previous demonstrations of quantum digital signatures assume trusted measurement devices. This renders them vulnerable against detector side-channel attacks, just like quantum key distribution. Here, we exploit a measurement-device-independent (MDI) quantum network, over a 200-square-kilometer metropolitan area, to perform a field test of a three-party measurement-device-independent quantum digital signature (MDI-QDS) scheme that is secure against any detector side-channel attack. In so doing, we are able to successfully sign a binary message with a security level of about $10^{-7}$. Remarkably, our work demonstrates the feasibility of MDI-QDS for practical applications.
\end{abstract}

\maketitle
Digital signatures are cryptographic schemes that are widely used to guarantee both the authenticity and the transferability of digital messages and documents. They play an essential role in many applications such as software distribution, financial transactions and e-mails. However, the security of currently used public-key digital signature schemes rely on computational assumptions, such as
the difficulty of factorizing large numbers~\cite{rivest1978method} or finding discrete logarithms~\cite{elgamal1984public}. Thus, advances in the development of efficient algorithms or a quantum computer can threaten their security.

Quantum digital signatures (QDS)~\cite{gottesman2001quantum}, on the other hand, can offer information-theoretic security based on quantum mechanics, given that the participants pre-share some secret keys for authentication purposes. That is, they guarantee no forging ({\it i.e.}, the message is signed by a legitimate sender and it has not been modified) and non-repudiation ({\it i.e.}, the sender cannot successfully deny the signature of the message) despite any future computational advance. This justifies the great attention that this topic has received recently. Indeed, QDS schemes based on coherent states~\cite{andersson2006experimentally, clarke2012experimental} and schemes that do not need the use of quantum memories~\cite{dunjko2014quantum, collins2014realization} have been proposed and experimentally demonstrated. Also, QDS protocols implementable with only {\blue QKD} components have been designed ~\cite{wallden2015quantum} and experimentally tested~\cite{donaldson2016experimental}. Remarkably, the need for trust on the quantum channels has also been removed~\cite{amiri2016secure, yin2016practical}. All these efforts have paved the way for the development of more practical QDS schemes~\cite{Croal:2016:Free,Yin:2016:exp,Collins:2016:exp}.

Despite this tremendous progress, however, in practice it is still very challenging to guarantee the security of the implementations. This is so because, just as for QKD, also here there is a big gap between practical realizations and the theoretical models that are assumed in the security proofs. As a result, we face security loopholes, or so-called side-channels, that could seriously threaten the security of QDS schemes. Indeed, detector side-channel attacks \cite{lydersen2010hacking,Zhao:2008:Quantum,weier2011quantum} are arguably the most important threat. Very recently, motivated by the concept of measurement-device-independent (MDI) QKD~\cite{lo2012measurement}, Puthoor $et~al.$~\cite{puthoor2016measurement} introduced a MDI-QDS scheme that is secure against all detector side-channel attacks.

In this Letter, we report the first experimental demonstration of a three-party MDI-QDS protocol which is immune to detector side-channel attacks and allows the signature of binary messages with a security level of $10^{-7}$. This implementation makes use of a MDI quantum network with a star topology that is deployed over a 200-square-kilometer metropolitan field. Our work demonstrates the feasibility of MDI-QDS schemes for practical applications.

In the MDI-QDS protocol of~\cite{puthoor2016measurement} there are at least three parties. One party (say for instance Alice) acts as a signer, while the other two parties (say Bob and Charlie) act as recipients. All parties are pairwise connected via authenticated classical channels. Also, they are connected to a relay (Eve) via quantum channels. The quantum channels between Bob and Eve, and Charlie and Eve, can be used to generate a secret key between Bob and Charlie by means of MDI-QKD. This secret key allows them to interchange messages in full secrecy by means of one-time pad encryption.

\begin{figure*}
\centering
\resizebox{15cm}{!}{\includegraphics{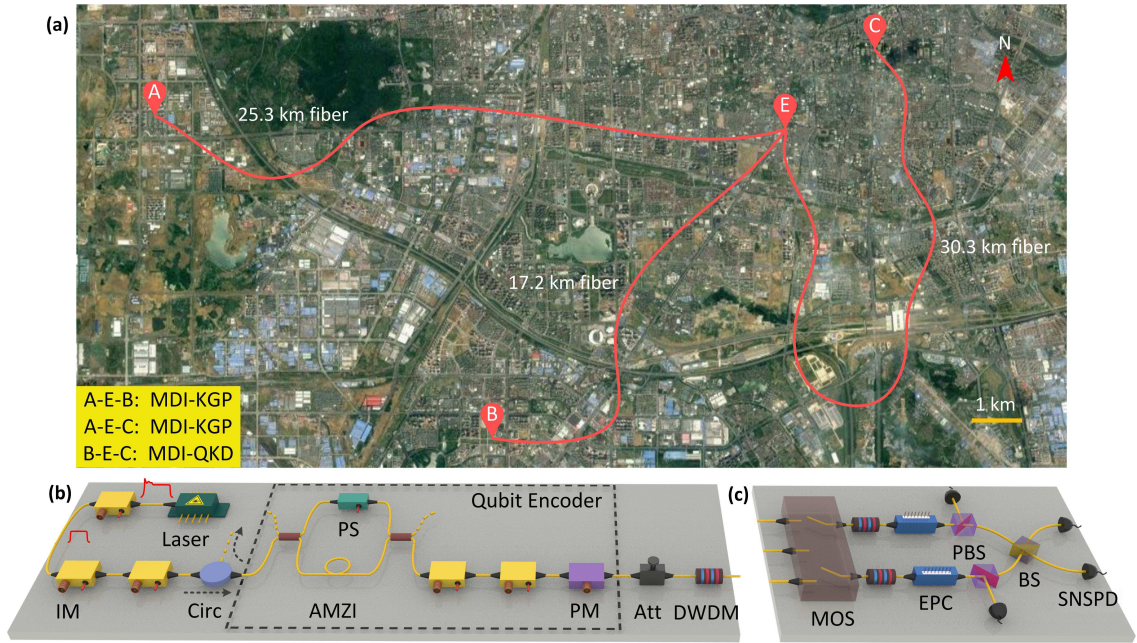}}
\caption{MDI-QDS experiment in a Hefei optical fiber network. (a) Birds-eye view of the MDI-QDS experiment.
Alice `A' is located in the Animation Industry Park (\textrm{N31$^{\circ}50'6.24''$}, \textrm{E117$^{\circ}7'52.08''$}), Bob `B' at the administrative committee of Hefei (\textrm{N31$^{\circ}47'4.56''$}, \textrm{E117$^{\circ}12'58.04''$}), Charlie `C' in an office building (\textrm{N31$^{\circ}50'56.84''$}, \textrm{E117$^{\circ}16'50.14''$}) and Eve `E' at the University of Science and Technology of China (\textrm{N31$^{\circ}50'7.56''$}, \textrm{E117$^{\circ}12'58.04''$}). The quantum link A-E-B (A-E-C) is used to perform the MDI-KGP, which generates correlated $L$-bit strings $A_{m}^{\rm B}$ and $K_{m}^{\rm B}$ ($A_{m}^{\rm C}$ and $K_{m}^{\rm C}$) between Alice and Bob (Charlie). The quantum link B-E-C is used to carry out the MDI-QKD scheme, which generates a secure key between Bob and Charlie. This key is used to one-time pad encrypt the information exchanged by these two users during the symmetrization step of the MDI-QDS scheme. (b) Alice's set-up. The set-ups of Bob and Charlie are identical to the one of Alice. The internally modulated laser generates phase-randomized coherent state signal pulses. The first intensity modulator (IM) removes the overshoot rising edge of the signal pulses. The following two IMs implement the decoy state method~\cite{hwang2003quantum,lo2005decoy,wang2005beating}. An asymmetrical Mach-Zehnder interferometer (AMZI) with a phase shifter (PS), in combination with two IMs and one phase modulator (PM), form a qubit encoder, which realizes a time-bin phase encoding. The attenuator (Att) is electrically controlled; it can quickly and automatically change the intensity of the outcoming signals to realize either the Hong-Ou-Mandel (HOM) interference or single-photon level preparation. Spurious emission is removed by means of a dense wavelength division multiplexor (DWDM). (c) Eve's set-up. An 8-by-4 mechanical optical switch (MOS) implements the routing function.  An electric polarization controller (EPC), a polarization beam splitter (PBS) and a superconducting nanowire single-photon detector (SNSPD) form the polarization feedback system. The beam splitter (BS) and two SNSPDs are used to implement the Bell state measurement (BSM).
}\label{Fig:NetDeploySetup}
\end{figure*}

The MDI-QDS protocol consists of two stages: the distribution stage and the messaging stage. Quantum communication is needed only in the former, where Alice uses a so-called MDI key generation protocol (MDI-KGP) to generate correlated $L$-bit strings $A_{\rm 0}^{\rm B},~A_{\rm 1}^{\rm B}$ and $A_{\rm 0}^{\rm C},~A_{\rm 1}^{\rm C}$ with Bob and Charlie, respectively. The corresponding strings held by Bob (Charlie) are denoted by $K_{m}^{\rm B}$ ($K_{m}^{\rm B}$), with $m=0,1$. Note that the strings $A_m^{\rm B}$ ($A_m^{\rm C}$) and $K_m^{\rm B}$ ($K_m^{\rm C}$) do not need to be identical, but they just need to be sufficiently correlated. The quantum stage of the MDI-KGP is equal to that of MDI-QKD, but its classical data post-processing stage is different because in the MDI-KGP there is no need to apply error correction and privacy amplification. After the MDI-KGP, Bob and Charlie symmetrize their strings. For this, say Bob randomly chooses half of the bits of each of $K_m^{\rm B}$ and sends them (as well as the information of the positions of the bits chosen) to Charlie using a secure channel. Similarly, Charlie does the same with $K_m^{\rm C}$. We denote Bob's (Charlie's) bit strings after the symmetrization step by $S_m^{\rm B}$ ($S_m^{\rm C}$).

Finally, in the messaging stage, which typically occurs much later and where only classical communication takes place, Alice can sign a binary message $m$ by simply sending $(m, \textrm{Sig}_m)$ to the desired recipient (say Bob), where the signature $\textrm{Sig}_m = \left( {A_m^{\rm B},{\kern 1pt} {\kern 1pt} A_m^{\rm C}{\kern 1pt} } \right)$. To verify that $m$ indeed comes from Alice, Bob checks whether $\textrm{Sig}_m$ matches his bit string $S_m^{\rm B}$. For this, he checks separately the part of $S_m^{\rm B}$ received directly from Alice and that received from Charlie, and he records the number of mismatches in each part. If the number of mismatches in both parts is below $s_a(L/2)$, where $s_a$ is a pre-fixed threshold value satisfying $0 < s_a < 1/2$, then Bob accepts the message as authentic. Otherwise, he rejects it. If Bob wants to demonstrate Charlie that Alice signed $m$, he sends him $(m,~Sig_m)$. Then Charlie performs a similar check to that done by Bob and only accepts $m$ if the number of mismatches in both halves of $S_m^{\rm C}$ is below $s_v(L/2)$, with $ 0 < s_a < s_v < 1/2$. In so doing, the MDI-QDS protocol is secure
against general forging and repudiation attacks~\cite{puthoor2016measurement}.

In order to experimentally demonstrate this MDI-QDS scheme we use the MDI quantum network that has been deployed in the city of Hefei, China. This metropolitan network has been recently used to successfully demonstrate MDI-QKD~\cite{tang2016measurement}. As shown in Fig. \ref{Fig:NetDeploySetup}(a), Alice, Bob and Charlie are connected to Eve, with a 25.3 km, 17.2 km and 30.3 km deployed single-mode optical fiber which has a propagation loss of 9.2 dB, 5.1 dB and 8.1 dB, respectively. In collaboration with Eve, Alice and Bob (Charlie) exploit the insecure quantum link A-E-B (A-E-C) to implement the MDI-KGP. Also, Bob and Charlie use the insecure quantum link B-E-C to implement the MDI-QKD protocol. For this, Eve's Bell state measurement (BSM) device is shared between Alice, Bob and Charlie. This is done by using an 8-by-4 mechanical optical switch (MOS) as a router, allowing us to perform three quantum protocols successively.

Since the quantum stage of the MDI-KGP is identical to that of MDI-QKD, identical state preparation setups are installed for the three participants Alice, Bob, and Charlie, who communicate with each other through classical channels and exchange quantum signals with Eve by means of quantum channels. This is illustrated in Figs.  \ref{Fig:NetDeploySetup}(b) and \ref{Fig:NetDeploySetup}(c). At each site, phase-randomized signal pulses at a repetition rate of 75 MHz are generated with an internally modulated distributed feedback laser. The wavelength of each signal pulse is 1550.12 nm and its pulse width is 2.5 ns. The intensities of the signal state, the decoy state and the vacuum state are $\mu=0.33$, $\nu=0.1$ and $w=0$, respectively. The corresponding probability distributions are set as 25.6\%, 58.4\% and 16\%, respectively. A time-bin phase encoding scheme~\cite{Ma:2012:Alternative} is used to prepare Bennett-Brassard 1984 states~\cite{bennett1984quantum}, where the delay between two time-bins is 6.37 ns. The signal (decoy) states are all prepared using the Z basis (the Z or the X bases with probability distribution 36.9\% and 63.1\%, respectively). In the case of the vacuum states $w$, it is not necessary to distinguish between the two bases. After applying a filter and a single-photon level modulation, each optical pulse is sent to Eve through the deployed fiber. A successful BSM result corresponds to coincidence counts in opposite time bins, which indicates a projection onto the singlet Bell state $\ket{\Psi^{-}}$. This means that the data shared between the participants is anti-correlated and one of them has to flip the bits to match those of the other participant. In the BSM, the efficiency of the time window for a single time-bin is about 90\%. The two superconducting nanowire single-photon detectors (SNSPDs) of the BSM work at 2.05 K and have detection efficiencies of 66\% and 64\%, respectively, as well as a dark count rate of 30 Hz. Also, the spurious noise of the deployed fiber brings dozens of extra dark counts per second. The inner insertion loss of Eve's system is 6.2 dB for the A-E-B link, 6.2 dB for the A-E-C link and 7 dB for the B-E-C link, respectively. This insertion loss includes the loss contribution from the MOS, the dense wavelength division multiplexor (DWDM), the electric polarization controller (EPC), the polarization beam splitter (PBS), the beam splitter (BS) and the optical fiber connection.

To achieve high-visibility two-photon interference in the BSM, the incoming photons have to be indistinguishable. For this, Eve uses three independent lasers at a wavelength of 1570 nm that generate 500 KHz signals to synchronize the entire system. Also, a programmable delay chip with 10 ps timing resolution is used to guarantee a precise overlap of the two interfering pulses~\cite{Yin:2016:MDIQKD}. The optical signal of the shared phase feedback laser with a wavelength of 1550.12 nm is divided into three beams by a BS. Each beam is sent to Alice, Bob and Charlie, respectively. The phase reference frame is stabilized by using a phase shifter (PS) and two power meters~\cite{tang2016measurement}. The synchronization signal and the phase feedback signal are multiplexed in an additional deployed fiber. The polarization reference frame is stabilized by using an EPC, a PBS, a SNSPD and a fast axis blocked polarization maintaining BS. Also, we use the HOM dip to calibrate the wavelength difference between the two interfering pulses~\cite{tang2016measurement}.

\begin{table}
\begin{ruledtabular}
\caption{\label{tab:Key} Parameters $n_0^{b,c}$, $n_1^{b,c}$, $e_1^{b,c}$, and $\textrm{leak}_{\textrm{EC}}^{b,c}$, with $b,c \in \{\nu, \mu\}$, for the MDI-QKD link between Bob and Charlie.}
\begin{tabular}{c|cccc}
&$\mu\mu$ & $\mu\nu$ & $\nu\mu$  & $\nu\nu$ \\
$n_{0}^{b,c}$ & {0} & {0} & {0} & {0}\\
$n_{1}^{b,c}$ & {13144467} &{4208999} &{4208978} &{1346138}\\
$e_{1}^{b,c}$ & {20.57\%} &{20.61\%} &{20.61\%} &{20.72\%}\\
$\textrm{leak}_{\textrm{EC}}^{ab}$ & 764378 & 446414 & 290251 & 133085\\
\end{tabular}
\end{ruledtabular}
\end{table}

We have run the MDI-KGP between the participants for 73423 (149987) seconds to accumulate data for the pair Alice and Bob (Alice and Charlie). Also, we accumulated data for 81630 seconds during the MDI-QKD session between Bob and Charlie. The experimental results are in the Supplemental Material. In the case of the MDI-QKD link between Bob and Charlie, we distill key from the Z basis data, while the X basis data is all used for parameter estimation. The length $\ell$ of the resulting secret key which guarantees that the MDI-QKD protocol is $\epsilon_{\textrm{QKD}}$-secure, {\it i.e.}, it is both $\epsilon_{\textrm{cor}}$-correct and $\epsilon_{\textrm{sec}}$-secret with $\epsilon_{\textrm{sec}}+\epsilon_{\textrm{cor}}\leq \epsilon_{\textrm{QKD}}$, is given by \cite{curty2014finite}
\begin{eqnarray}\label{eq_key}
\ell= \sum\limits_{b,c \in \{0,\;\nu ,\;\mu \} } n_0^{b,c} + n_1^{b,c}\left[ {1 - h\left( {e_1^{b,c}} \right)} \right]- \textrm{leak}_{\textrm{EC}}^{b,c}\nonumber 
 - {\log _2}\frac{8}{{\epsilon_{\textrm{cor}}^{b,c}}}- 2{\log _2}\frac{2}{{\varepsilon {'^{b,c}}{{\hat \varepsilon }^{b,c}}}} - 2{\log _2}\frac{1}{{2\varepsilon _{\textrm{PA}}^{b,c}}},
\end{eqnarray}
where $\epsilon_{\textrm{cor}}=\sum_{b,c} {\epsilon _{\textrm{cor} }^{b,c}}$ and $\epsilon_{\textrm{sec}}=\sum_{b,c} {\epsilon _{\textrm{sec} }^{b,c}}$, with $\epsilon _{\textrm{sec} }^{b,c} = 2\left( {{\varepsilon{'^{b,c}}} + 2\varepsilon _e^{b,c} + {{\hat \varepsilon }^{b,c}}} \right) + \varepsilon _\beta ^{b,c} + \varepsilon _0^{b,c} + \varepsilon _1^{b,c} + \varepsilon _{\textrm{PA}}^{b,c}$. The parameters $\varepsilon _0^{b,c}$, $\varepsilon _1^{b,c}$ and $\varepsilon _e^{b,c}$ denote the failure probability associated with the estimation of $n_0^{b,c}$, $n_1^{b,c}$ and $e_1^{b,c}$, respectively. $\epsilon _{\textrm{cor} }^{b,c}$ and $\varepsilon _{\textrm{PA}}^{b,c}$ represent the failure probability of the error verification and the privacy amplification steps, respectively. See~\cite{curty2014finite} for further details. Here, we use Bob's data as the reference raw key. Therefore, in Eq.~(\ref{eq_key}), $n_0^{b,c}$ ($n_1^{b,c}$) is a lower bound for the number of events where Bob (Bob and Charlie) emitted a vacuum (single-photon) state that produced a successful BSM result, given that Bob and Charlie selected the intensity settings $b$ and $c$, with $b,c \in \{0,\nu, \mu\}$, respectively. $e_1^{b,c}$ is an upper bound for the single-photon phase-error rate, and $\textrm{leak}_{\textrm{EC}}^{b,c}$ is the information revealed during the error correction step with $h\left( x \right) =  - x{\log _2}\left( x \right) - \left( {1 - x} \right){\log _2}\left( {1 - x} \right)$ being the Shannon entropy function.

\begin{table*}
\begin{ruledtabular}
\caption{\label{tab:Key} The value of the different parameters in the MDI-QDS experiment.}
\begin{tabular}{cccccccccc}
$\bar{E}$ &$s_{a}$ &$s_{a}L/2$ & $s_{v}$ &$s_{v}L/2$ & $p_{E}$ & $\varepsilon_{\textrm{rob}}$ & $\varepsilon_{\textrm{rep}}$ & $\varepsilon_{\textrm{for}}$ \\
{0.25\%} & {0.27\%} & {1073} &{1.21\%} & {4748} & {1.23\%} &{$2\times10^{-8}$} & {$1.51\times10^{-7}$} &{$9.76\times10^{-8}$} \\
\end{tabular}
\end{ruledtabular}
\end{table*}

According to Eq. (\ref{eq_key}), in principle one can distill secret key from all the possible combinations of the intensity settings. In our experiment, however, we find that only the data corresponding to the intensity settings $b,c \in \{\nu, \mu\}$ provide a positive key rate. The values of the parameters $n_0^{b,c}$, $n_1^{b,c}$, $e_1^{b,c}$, and $\textrm{leak}_{\textrm{EC}}^{b,c}$, with $b,c \in \{\nu, \mu\}$, are shown in Table I. Also, we use the Cascade algorithm to implement error correction \cite{brassard1993secret}, and a Toeplitz matrix to perform privacy amplification. The random bit string that is needed to generate the Toeplitz matrix is obtained in a previous QKD experiment. The security level of the MDI-QKD protocol is set as $\epsilon_{\textrm{QKD}}=8\times10^{-8}$ and we obtain a $\epsilon_{\textrm{QKD}}$-secure key of length $\ell=4724819$ bits.

In the symmetrization step of the MDI-QDS scheme, the position information about the exchanged bits is encoded as follows. For each $L$-bit string $K_m^{\rm B}$ ($K_m^{\rm C}$) we prepare an $L$-bit string whose elements are set to $0$ or $1$ depending on whether or not the equivalent element of $K_m^{\rm B}$ is sent to Charlie (Bob). That is, for each $K_m^{\rm B}$ (or $K_m^{\rm C}$) we need $3L/2$ secret bits for one-time pad encryption ($L/2$ bits are used to encrypt the actual bits exchanged between the participants and $L$ bits are used to encrypt the string with the position information). In total we need $4\times3L/2=6L$ secret bits, and thus we select $6L\leq\ell$. For our experiment, we choose $L=787468$.

In the MDI-KGP between Alice and Bob (Charlie), the signature bit strings $A_{m}^{\rm B}$ ($A_{m}^{\rm C}$) are generated only from the data associated with those events where both Alice and Bob (Charlie) use the Z basis and the signal intensity $\mu$. Moreover, Alice and Bob (Charlie) split the correlated bit strings generated in one run of the MDI-KGP into two equally long bit strings. Then, each of Alice and Bob (Charlie), selects $L$ bits at random to form the bit strings $A_{0}^{\rm B}$ and $A_{1}^{\rm B}$ ($A_{0}^{\rm C}$ and $A_{1}^{\rm C}$), respectively. The remaining bits are all announced to estimate the bit error rate of that string. The results associated with the randomly selected signatures are in the Supplemental Material. With this bit error rate information, we use Serfling inequality~\cite{serfling1974probability} to estimate an upper bound for the error rate between the part of the string $K_{m}^{B}$ ($K_{m}^{C}$) that Bob (Charlie) keeps for himself and $A_{m}^{\rm B}$ ($A_{m}^{\rm C}$), which is true except for a minuscule probability $\varepsilon_{\textrm{PE}}$. We denote these upper bounds by $E_{m}^{\rm B}$ and $E_{m}^{\rm C}$, respectively, and we set $\bar{E}=\max\{E_{m}^{\rm B},E_{m}^{\rm C}\}$.

Finally, to evaluate the security of the MDI-QDS experiment, we follow the procedure introduced in~\cite{puthoor2016measurement}. This involves the calculation of the minimum rate, $p_E$, at which Eve is likely to make errors when guessing the part of $K_{m}^{B}$ that Bob keeps for himself. Also, one has to select certain parameters $s_{a}$ and $s_{v}$ such that $\bar{E}<s_{a}<s_{v}<p_{E}$ to guarantee security against repudiation and forging. As a result, we have that the probability $\varepsilon_{\textrm{rep}}$ of successful repudiation, {\it i.e.}, that Alice can make Bob accept a message $m$ and Charlie rejects it when it is transferred to him, is~\cite{puthoor2016measurement}
\begin{equation} \label{eq1}
\varepsilon_{\rm rep} \le 2\exp \left[ { - \frac{1}{4}{{\left( {{s_v} - {s_a}} \right)}^2{L}}} \right]+\epsilon_{\rm QKD}.
\end{equation}
The first term on the RHS of this equation corresponds to the probability of success repudiation given that Bob and Charlie share a perfectly secure secret key before they perform the MDI-KGP~\cite{puthoor2016measurement}, while the second term takes into account the probability that the secret key delivered by the MDI-QKD protocol is not secure. Similarly, the probability $\varepsilon_{\textrm{for}}$ of successful forging, {\it i.e.}, that Bob can generate a fraudulent declaration $(m,\textrm{Sig}_{m})$ that Charlie accepts, satisfies~\cite{puthoor2016measurement}
\begin{equation} \label{eq1}
\begin{aligned}
\varepsilon_{\textrm{for}} \le \frac{1}{f}\left(2^{-\frac{L}{2}[h(p_{E})-h(s_{v})]}+\varepsilon\right) + f + {\varepsilon _{{\textrm{PE}}}} + {\varepsilon _{\rm est}},
\end{aligned}
\end{equation}
where the parameters $\varepsilon$, $\varepsilon _{\rm est}$ and $f$ are related to the failure probability when estimating $p_E$, and $\varepsilon _{{\textrm{PE}}}$ is related to the robustness $\varepsilon_{\textrm{rob}}$ of the protocol. See Supplemental Material for more details. The value of each of these parameters in the MDI-QDS experiment is shown in Table II.

After performing the two MDI-KGPs and the MDI-QKD scheme to generate the correlated bit strings $A_{m}^{\rm B}$, $K_{m}^{\rm B}$, $A_{m}^{\rm C}$ and $K_{m}^{\rm C}$, as well as a secret key of length $\ell$, we also implemented experimentally the classical network that is needed to actually sign a binary message. This includes the implementation of the symmetrization step to generate the bit strings $S_m^{\rm B}$ and $S_m^{\rm C}$, and the realization of the messaging stage. All the random bit strings needed for random sampling as well as the secret key that is used to authenticate the classical communications in the MDI-QDS experiment are taken from previous QKD experiments. The secret key generated in the MDI-QKD link is employed to one-time pad encrypt the information exchanged in the symmetrization step. In this work, Alice decides to sign the message $m=1$ and sends $(1,\textrm{Sig}_{1})$ to Bob in the messaging stage. Bob calculates the number of mismatches, 897, between $A_{\rm 1}^{\rm B}$ and the part of $K_{\rm 1}^{\rm B}$ that he keeps for himself, and, 508, between $A_{\rm 1}^{\rm C}$ and the part of $K_{\rm 1}^{\rm C}$ received from Charlie. He accepts the message and forwards $(1,\textrm{Sig}_{{\rm 1}})$ to Charlie since both mismatches are below their corresponding threshold. Charlie performs a similar check like Bob and accepts $m$ because the number of mismatches, 502, between $A_{\rm 1}^{\rm C}$ and the part of $K_{\rm 1}^{\rm C}$ that he keeps for himself, and, 914, between $A_{\rm 1}^{\rm B}$ and the part of $K_{\rm 1}^{\rm B}$ received from Bob are below their corresponding thresholds.

In conclusion, we have experimentally demonstrated for the first time a complete MDI-QDS protocol in a field test with a failure probability about $10^{-7}$. This scheme is information-theoretically secure and is free of any detector side-channel. In so doing, we have successfully signed a binary message between three parties. We remark that the signature efficiency of this work is relatively low because we did not perform the full parameter optimization. As in the case of MDI-QKD, we believe that the use of the four-intensity decoy-state method~\cite{Yin:2016:MDIQKD} and increasing the system clock rate~\cite{comandar2016quantum} would permit us to significantly increase the signature efficiency.

This work has been supported by the National Fundamental Research Program (under Grant No. 2013CB336800), the National Natural Science Foundation of China, the Chinese Academy of Science and the Science Fund of Anhui Province for Outstanding Youth. M.C. gratefully acknowledges support from the Galician Regional Government (Grant No. EM2014/033, and consolidation of Research Units:AtlantTIC), MINECO, the Fondo Europeo de Desarrollo Regional (FEDER) through Grant No. TEC2014-54898-R, and the European Commission (Project QCALL). EA and IVP acknowledge support from EPSRC grant EP/M013472/1. W.W. gratefully acknowledges support from the National Natural Science Foundation of China under Grant No. 61472446.



\begin{thebibliography}{30}%
\makeatletter
\providecommand \@ifxundefined [1]{%
 \@ifx{#1\undefined}
}%
\providecommand \@ifnum [1]{%
 \ifnum #1\expandafter \@firstoftwo
 \else \expandafter \@secondoftwo
 \fi
}%
\providecommand \@ifx [1]{%
 \ifx #1\expandafter \@firstoftwo
 \else \expandafter \@secondoftwo
 \fi
}%
\providecommand \natexlab [1]{#1}%
\providecommand \enquote  [1]{``#1''}%
\providecommand \bibnamefont  [1]{#1}%
\providecommand \bibfnamefont [1]{#1}%
\providecommand \citenamefont [1]{#1}%
\providecommand \href@noop [0]{\@secondoftwo}%
\providecommand \href [0]{\begingroup \@sanitize@url \@href}%
\providecommand \@href[1]{\@@startlink{#1}\@@href}%
\providecommand \@@href[1]{\endgroup#1\@@endlink}%
\providecommand \@sanitize@url [0]{\catcode `\\12\catcode `\$12\catcode
  `\&12\catcode `\#12\catcode `\^12\catcode `\_12\catcode `\%12\relax}%
\providecommand \@@startlink[1]{}%
\providecommand \@@endlink[0]{}%
\providecommand \url  [0]{\begingroup\@sanitize@url \@url }%
\providecommand \@url [1]{\endgroup\@href {#1}{\urlprefix }}%
\providecommand \urlprefix  [0]{URL }%
\providecommand \Eprint [0]{\href }%
\providecommand \doibase [0]{http://dx.doi.org/}%
\providecommand \selectlanguage [0]{\@gobble}%
\providecommand \bibinfo  [0]{\@secondoftwo}%
\providecommand \bibfield  [0]{\@secondoftwo}%
\providecommand \translation [1]{[#1]}%
\providecommand \BibitemOpen [0]{}%
\providecommand \bibitemStop [0]{}%
\providecommand \bibitemNoStop [0]{.\EOS\space}%
\providecommand \EOS [0]{\spacefactor3000\relax}%
\providecommand \BibitemShut  [1]{\csname bibitem#1\endcsname}%
\let\auto@bib@innerbib\@empty
\bibitem [{\citenamefont {Rivest}\ \emph {et~al.}(1978)\citenamefont {Rivest},
  \citenamefont {Shamir},\ and\ \citenamefont {Adleman}}]{rivest1978method}%
  \BibitemOpen
  \bibfield  {author} {\bibinfo {author} {\bibfnamefont {R.~L.}\ \bibnamefont
  {Rivest}}, \bibinfo {author} {\bibfnamefont {A.}~\bibnamefont {Shamir}}, \
  and\ \bibinfo {author} {\bibfnamefont {L.}~\bibnamefont {Adleman}},\
  }\href@noop {} {\bibfield  {journal} {\bibinfo  {journal} {Communications of
  the ACM}\ }\textbf {\bibinfo {volume} {21}},\ \bibinfo {pages} {120}
  (\bibinfo {year} {1978})}\BibitemShut {NoStop}%
\bibitem [{\citenamefont {ElGamal}(1984)}]{elgamal1984public}%
  \BibitemOpen
  \bibfield  {author} {\bibinfo {author} {\bibfnamefont {T.}~\bibnamefont
  {ElGamal}},\ }in\ \href@noop {} {\emph {\bibinfo {booktitle} {Workshop on the
  Theory and Application of Cryptographic Techniques}}}\ (\bibinfo
  {organization} {Springer},\ \bibinfo {year} {1984})\ pp.\ \bibinfo {pages}
  {10--18}\BibitemShut {NoStop}%
\bibitem [{\citenamefont {Gottesman}\ and\ \citenamefont
  {Chuang}(2001)}]{gottesman2001quantum}%
  \BibitemOpen
  \bibfield  {author} {\bibinfo {author} {\bibfnamefont {D.}~\bibnamefont
  {Gottesman}}\ and\ \bibinfo {author} {\bibfnamefont {I.}~\bibnamefont
  {Chuang}},\ }\href@noop {} {\bibfield  {journal} {\bibinfo  {journal} {arXiv
  preprint quant-ph/0105032}\ } (\bibinfo {year} {2001})}\BibitemShut {NoStop}%
\bibitem [{\citenamefont {Andersson}\ \emph {et~al.}(2006)\citenamefont
  {Andersson}, \citenamefont {Curty},\ and\ \citenamefont
  {Jex}}]{andersson2006experimentally}%
  \BibitemOpen
  \bibfield  {author} {\bibinfo {author} {\bibfnamefont {E.}~\bibnamefont
  {Andersson}}, \bibinfo {author} {\bibfnamefont {M.}~\bibnamefont {Curty}}, \
  and\ \bibinfo {author} {\bibfnamefont {I.}~\bibnamefont {Jex}},\ }\href@noop
  {} {\bibfield  {journal} {\bibinfo  {journal} {Physical Review A}\ }\textbf
  {\bibinfo {volume} {74}},\ \bibinfo {pages} {022304} (\bibinfo {year}
  {2006})}\BibitemShut {NoStop}%
\bibitem [{\citenamefont {Clarke}\ \emph {et~al.}(2012)\citenamefont {Clarke},
  \citenamefont {Collins}, \citenamefont {Dunjko}, \citenamefont {Andersson},
  \citenamefont {Jeffers},\ and\ \citenamefont
  {Buller}}]{clarke2012experimental}%
  \BibitemOpen
  \bibfield  {author} {\bibinfo {author} {\bibfnamefont {P.~J.}\ \bibnamefont
  {Clarke}}, \bibinfo {author} {\bibfnamefont {R.~J.}\ \bibnamefont {Collins}},
  \bibinfo {author} {\bibfnamefont {V.}~\bibnamefont {Dunjko}}, \bibinfo
  {author} {\bibfnamefont {E.}~\bibnamefont {Andersson}}, \bibinfo {author}
  {\bibfnamefont {J.}~\bibnamefont {Jeffers}}, \ and\ \bibinfo {author}
  {\bibfnamefont {G.~S.}\ \bibnamefont {Buller}},\ }\href@noop {} {\bibfield
  {journal} {\bibinfo  {journal} {Nature communications}\ }\textbf {\bibinfo
  {volume} {3}},\ \bibinfo {pages} {1174} (\bibinfo {year} {2012})}\BibitemShut
  {NoStop}%
\bibitem [{\citenamefont {Dunjko}\ \emph {et~al.}(2014)\citenamefont {Dunjko},
  \citenamefont {Wallden},\ and\ \citenamefont
  {Andersson}}]{dunjko2014quantum}%
  \BibitemOpen
  \bibfield  {author} {\bibinfo {author} {\bibfnamefont {V.}~\bibnamefont
  {Dunjko}}, \bibinfo {author} {\bibfnamefont {P.}~\bibnamefont {Wallden}}, \
  and\ \bibinfo {author} {\bibfnamefont {E.}~\bibnamefont {Andersson}},\
  }\href@noop {} {\bibfield  {journal} {\bibinfo  {journal} {Physical review
  letters}\ }\textbf {\bibinfo {volume} {112}},\ \bibinfo {pages} {040502}
  (\bibinfo {year} {2014})}\BibitemShut {NoStop}%
\bibitem [{\citenamefont {Collins}\ \emph {et~al.}(2014)\citenamefont
  {Collins}, \citenamefont {Donaldson}, \citenamefont {Dunjko}, \citenamefont
  {Wallden}, \citenamefont {Clarke}, \citenamefont {Andersson}, \citenamefont
  {Jeffers},\ and\ \citenamefont {Buller}}]{collins2014realization}%
  \BibitemOpen
  \bibfield  {author} {\bibinfo {author} {\bibfnamefont {R.~J.}\ \bibnamefont
  {Collins}}, \bibinfo {author} {\bibfnamefont {R.~J.}\ \bibnamefont
  {Donaldson}}, \bibinfo {author} {\bibfnamefont {V.}~\bibnamefont {Dunjko}},
  \bibinfo {author} {\bibfnamefont {P.}~\bibnamefont {Wallden}}, \bibinfo
  {author} {\bibfnamefont {P.~J.}\ \bibnamefont {Clarke}}, \bibinfo {author}
  {\bibfnamefont {E.}~\bibnamefont {Andersson}}, \bibinfo {author}
  {\bibfnamefont {J.}~\bibnamefont {Jeffers}}, \ and\ \bibinfo {author}
  {\bibfnamefont {G.~S.}\ \bibnamefont {Buller}},\ }\href@noop {} {\bibfield
  {journal} {\bibinfo  {journal} {Physical review letters}\ }\textbf {\bibinfo
  {volume} {113}},\ \bibinfo {pages} {040502} (\bibinfo {year}
  {2014})}\BibitemShut {NoStop}%
\bibitem [{\citenamefont {Wallden}\ \emph {et~al.}(2015)\citenamefont
  {Wallden}, \citenamefont {Dunjko}, \citenamefont {Kent},\ and\ \citenamefont
  {Andersson}}]{wallden2015quantum}%
  \BibitemOpen
  \bibfield  {author} {\bibinfo {author} {\bibfnamefont {P.}~\bibnamefont
  {Wallden}}, \bibinfo {author} {\bibfnamefont {V.}~\bibnamefont {Dunjko}},
  \bibinfo {author} {\bibfnamefont {A.}~\bibnamefont {Kent}}, \ and\ \bibinfo
  {author} {\bibfnamefont {E.}~\bibnamefont {Andersson}},\ }\href@noop {}
  {\bibfield  {journal} {\bibinfo  {journal} {Physical Review A}\ }\textbf
  {\bibinfo {volume} {91}},\ \bibinfo {pages} {042304} (\bibinfo {year}
  {2015})}\BibitemShut {NoStop}%
\bibitem [{\citenamefont {Donaldson}\ \emph {et~al.}(2016)\citenamefont
  {Donaldson}, \citenamefont {Collins}, \citenamefont {Kleczkowska},
  \citenamefont {Amiri}, \citenamefont {Wallden}, \citenamefont {Dunjko},
  \citenamefont {Jeffers}, \citenamefont {Andersson},\ and\ \citenamefont
  {Buller}}]{donaldson2016experimental}%
  \BibitemOpen
  \bibfield  {author} {\bibinfo {author} {\bibfnamefont {R.~J.}\ \bibnamefont
  {Donaldson}}, \bibinfo {author} {\bibfnamefont {R.~J.}\ \bibnamefont
  {Collins}}, \bibinfo {author} {\bibfnamefont {K.}~\bibnamefont
  {Kleczkowska}}, \bibinfo {author} {\bibfnamefont {R.}~\bibnamefont {Amiri}},
  \bibinfo {author} {\bibfnamefont {P.}~\bibnamefont {Wallden}}, \bibinfo
  {author} {\bibfnamefont {V.}~\bibnamefont {Dunjko}}, \bibinfo {author}
  {\bibfnamefont {J.}~\bibnamefont {Jeffers}}, \bibinfo {author} {\bibfnamefont
  {E.}~\bibnamefont {Andersson}}, \ and\ \bibinfo {author} {\bibfnamefont
  {G.~S.}\ \bibnamefont {Buller}},\ }\href@noop {} {\bibfield  {journal}
  {\bibinfo  {journal} {Physical Review A}\ }\textbf {\bibinfo {volume} {93}},\
  \bibinfo {pages} {012329} (\bibinfo {year} {2016})}\BibitemShut {NoStop}%
\bibitem [{\citenamefont {Amiri}\ \emph {et~al.}(2016)\citenamefont {Amiri},
  \citenamefont {Wallden}, \citenamefont {Kent},\ and\ \citenamefont
  {Andersson}}]{amiri2016secure}%
  \BibitemOpen
  \bibfield  {author} {\bibinfo {author} {\bibfnamefont {R.}~\bibnamefont
  {Amiri}}, \bibinfo {author} {\bibfnamefont {P.}~\bibnamefont {Wallden}},
  \bibinfo {author} {\bibfnamefont {A.}~\bibnamefont {Kent}}, \ and\ \bibinfo
  {author} {\bibfnamefont {E.}~\bibnamefont {Andersson}},\ }\href@noop {}
  {\bibfield  {journal} {\bibinfo  {journal} {Physical Review A}\ }\textbf
  {\bibinfo {volume} {93}},\ \bibinfo {pages} {032325} (\bibinfo {year}
  {2016})}\BibitemShut {NoStop}%
\bibitem [{\citenamefont {Yin}\ \emph {et~al.}(2016{\natexlab{a}})\citenamefont
  {Yin}, \citenamefont {Fu},\ and\ \citenamefont {Chen}}]{yin2016practical}%
  \BibitemOpen
  \bibfield  {author} {\bibinfo {author} {\bibfnamefont {H.-L.}\ \bibnamefont
  {Yin}}, \bibinfo {author} {\bibfnamefont {Y.}~\bibnamefont {Fu}}, \ and\
  \bibinfo {author} {\bibfnamefont {Z.-B.}\ \bibnamefont {Chen}},\ }\href@noop
  {} {\bibfield  {journal} {\bibinfo  {journal} {Physical Review A}\ }\textbf
  {\bibinfo {volume} {93}},\ \bibinfo {pages} {032316} (\bibinfo {year}
  {2016}{\natexlab{a}})}\BibitemShut {NoStop}%
\bibitem [{\citenamefont {Croal}\ \emph {et~al.}(2016)\citenamefont {Croal},
  \citenamefont {Peuntinger}, \citenamefont {Heim}, \citenamefont {Khan},
  \citenamefont {Marquardt}, \citenamefont {Leuchs}, \citenamefont {Wallden},
  \citenamefont {Andersson},\ and\ \citenamefont
  {Korolkova}}]{Croal:2016:Free}%
  \BibitemOpen
  \bibfield  {author} {\bibinfo {author} {\bibfnamefont {C.}~\bibnamefont
  {Croal}}, \bibinfo {author} {\bibfnamefont {C.}~\bibnamefont {Peuntinger}},
  \bibinfo {author} {\bibfnamefont {B.}~\bibnamefont {Heim}}, \bibinfo {author}
  {\bibfnamefont {I.}~\bibnamefont {Khan}}, \bibinfo {author} {\bibfnamefont
  {C.}~\bibnamefont {Marquardt}}, \bibinfo {author} {\bibfnamefont
  {G.}~\bibnamefont {Leuchs}}, \bibinfo {author} {\bibfnamefont
  {P.}~\bibnamefont {Wallden}}, \bibinfo {author} {\bibfnamefont
  {E.}~\bibnamefont {Andersson}}, \ and\ \bibinfo {author} {\bibfnamefont
  {N.}~\bibnamefont {Korolkova}},\ }\href@noop {} {\bibfield  {journal}
  {\bibinfo  {journal} {Phys. Rev. Lett.}\ }\textbf {\bibinfo {volume} {117}},\
  \bibinfo {pages} {100503} (\bibinfo {year} {2016})}\BibitemShut {NoStop}%
\bibitem [{\citenamefont {Yin}\ \emph {et~al.}(2016{\natexlab{b}})\citenamefont
  {Yin}, \citenamefont {Fu}, \citenamefont {Liu}, \citenamefont {Tang},
  \citenamefont {Wang}, \citenamefont {You}, \citenamefont {Zhang},
  \citenamefont {Chen}, \citenamefont {Wang}, \citenamefont {Zhang},
  \citenamefont {Chen}, \citenamefont {Chen},\ and\ \citenamefont
  {Pan}}]{Yin:2016:exp}%
  \BibitemOpen
  \bibfield  {author} {\bibinfo {author} {\bibfnamefont {H.-L.}\ \bibnamefont
  {Yin}}, \bibinfo {author} {\bibfnamefont {Y.}~\bibnamefont {Fu}}, \bibinfo
  {author} {\bibfnamefont {H.}~\bibnamefont {Liu}}, \bibinfo {author}
  {\bibfnamefont {Q.-J.}\ \bibnamefont {Tang}}, \bibinfo {author}
  {\bibfnamefont {J.}~\bibnamefont {Wang}}, \bibinfo {author} {\bibfnamefont
  {L.-X.}\ \bibnamefont {You}}, \bibinfo {author} {\bibfnamefont {W.-J.}\
  \bibnamefont {Zhang}}, \bibinfo {author} {\bibfnamefont {S.-J.}\ \bibnamefont
  {Chen}}, \bibinfo {author} {\bibfnamefont {Z.}~\bibnamefont {Wang}}, \bibinfo
  {author} {\bibfnamefont {Q.}~\bibnamefont {Zhang}}, \bibinfo {author}
  {\bibfnamefont {T.-Y.}\ \bibnamefont {Chen}}, \bibinfo {author}
  {\bibfnamefont {Z.-B.}\ \bibnamefont {Chen}}, \ and\ \bibinfo {author}
  {\bibfnamefont {J.-W.}\ \bibnamefont {Pan}},\ }\href@noop {} {\bibfield
  {journal} {\bibinfo  {journal} {arXiv:1608.01086}\ } (\bibinfo {year}
  {2016}{\natexlab{b}})}\BibitemShut {NoStop}%
\bibitem [{\citenamefont {Collins}\ \emph {et~al.}(2016)\citenamefont
  {Collins}, \citenamefont {Amiri}, \citenamefont {Fujiwara}, \citenamefont
  {Honjo}, \citenamefont {Shimizu}, \citenamefont {Tamaki}, \citenamefont
  {Takeoka}, \citenamefont {Andersson}, \citenamefont {Buller},\ and\
  \citenamefont {Sasaki}}]{Collins:2016:exp}%
  \BibitemOpen
  \bibfield  {author} {\bibinfo {author} {\bibfnamefont {R.~J.}\ \bibnamefont
  {Collins}}, \bibinfo {author} {\bibfnamefont {R.}~\bibnamefont {Amiri}},
  \bibinfo {author} {\bibfnamefont {M.}~\bibnamefont {Fujiwara}}, \bibinfo
  {author} {\bibfnamefont {T.}~\bibnamefont {Honjo}}, \bibinfo {author}
  {\bibfnamefont {K.}~\bibnamefont {Shimizu}}, \bibinfo {author} {\bibfnamefont
  {K.}~\bibnamefont {Tamaki}}, \bibinfo {author} {\bibfnamefont
  {M.}~\bibnamefont {Takeoka}}, \bibinfo {author} {\bibfnamefont
  {E.}~\bibnamefont {Andersson}}, \bibinfo {author} {\bibfnamefont {G.~S.}\
  \bibnamefont {Buller}}, \ and\ \bibinfo {author} {\bibfnamefont
  {M.}~\bibnamefont {Sasaki}},\ }\href@noop {} {\bibfield  {journal} {\bibinfo
  {journal} {Opt. Lett.}\ }\textbf {\bibinfo {volume} {41}},\ \bibinfo {pages}
  {4883} (\bibinfo {year} {2016})}\BibitemShut {NoStop}%
\bibitem [{\citenamefont {Lydersen}\ \emph {et~al.}(2010)\citenamefont
  {Lydersen}, \citenamefont {Wiechers}, \citenamefont {Wittmann}, \citenamefont
  {Elser}, \citenamefont {Skaar},\ and\ \citenamefont
  {Makarov}}]{lydersen2010hacking}%
  \BibitemOpen
  \bibfield  {author} {\bibinfo {author} {\bibfnamefont {L.}~\bibnamefont
  {Lydersen}}, \bibinfo {author} {\bibfnamefont {C.}~\bibnamefont {Wiechers}},
  \bibinfo {author} {\bibfnamefont {C.}~\bibnamefont {Wittmann}}, \bibinfo
  {author} {\bibfnamefont {D.}~\bibnamefont {Elser}}, \bibinfo {author}
  {\bibfnamefont {J.}~\bibnamefont {Skaar}}, \ and\ \bibinfo {author}
  {\bibfnamefont {V.}~\bibnamefont {Makarov}},\ }\href@noop {} {\bibfield
  {journal} {\bibinfo  {journal} {Nature photonics}\ }\textbf {\bibinfo
  {volume} {4}},\ \bibinfo {pages} {686} (\bibinfo {year} {2010})}\BibitemShut
  {NoStop}%
\bibitem [{\citenamefont {Zhao}\ \emph {et~al.}(2008)\citenamefont {Zhao},
  \citenamefont {Fung}, \citenamefont {Qi}, \citenamefont {Chen},\ and\
  \citenamefont {Lo}}]{Zhao:2008:Quantum}%
  \BibitemOpen
  \bibfield  {author} {\bibinfo {author} {\bibfnamefont {Y.}~\bibnamefont
  {Zhao}}, \bibinfo {author} {\bibfnamefont {C.-H.~F.}\ \bibnamefont {Fung}},
  \bibinfo {author} {\bibfnamefont {B.}~\bibnamefont {Qi}}, \bibinfo {author}
  {\bibfnamefont {C.}~\bibnamefont {Chen}}, \ and\ \bibinfo {author}
  {\bibfnamefont {H.-K.}\ \bibnamefont {Lo}},\ }\href@noop {} {\bibfield
  {journal} {\bibinfo  {journal} {Phys. Rev. A}\ }\textbf {\bibinfo {volume}
  {78}},\ \bibinfo {pages} {042333} (\bibinfo {year} {2008})}\BibitemShut
  {NoStop}%
\bibitem [{\citenamefont {Weier}\ \emph {et~al.}(2011)\citenamefont {Weier},
  \citenamefont {Krauss}, \citenamefont {Rau}, \citenamefont {F{\"u}rst},
  \citenamefont {Nauerth},\ and\ \citenamefont
  {Weinfurter}}]{weier2011quantum}%
  \BibitemOpen
  \bibfield  {author} {\bibinfo {author} {\bibfnamefont {H.}~\bibnamefont
  {Weier}}, \bibinfo {author} {\bibfnamefont {H.}~\bibnamefont {Krauss}},
  \bibinfo {author} {\bibfnamefont {M.}~\bibnamefont {Rau}}, \bibinfo {author}
  {\bibfnamefont {M.}~\bibnamefont {F{\"u}rst}}, \bibinfo {author}
  {\bibfnamefont {S.}~\bibnamefont {Nauerth}}, \ and\ \bibinfo {author}
  {\bibfnamefont {H.}~\bibnamefont {Weinfurter}},\ }\href@noop {} {\bibfield
  {journal} {\bibinfo  {journal} {New Journal of Physics}\ }\textbf {\bibinfo
  {volume} {13}},\ \bibinfo {pages} {073024} (\bibinfo {year}
  {2011})}\BibitemShut {NoStop}%
\bibitem [{\citenamefont {Lo}\ \emph {et~al.}(2012)\citenamefont {Lo},
  \citenamefont {Curty},\ and\ \citenamefont {Qi}}]{lo2012measurement}%
  \BibitemOpen
  \bibfield  {author} {\bibinfo {author} {\bibfnamefont {H.-K.}\ \bibnamefont
  {Lo}}, \bibinfo {author} {\bibfnamefont {M.}~\bibnamefont {Curty}}, \ and\
  \bibinfo {author} {\bibfnamefont {B.}~\bibnamefont {Qi}},\ }\href@noop {}
  {\bibfield  {journal} {\bibinfo  {journal} {Physical Review Letters}\
  }\textbf {\bibinfo {volume} {108}},\ \bibinfo {pages} {130503} (\bibinfo
  {year} {2012})}\BibitemShut {NoStop}%
\bibitem [{\citenamefont {Puthoor}\ \emph {et~al.}(2016)\citenamefont
  {Puthoor}, \citenamefont {Amiri}, \citenamefont {Wallden}, \citenamefont
  {Curty},\ and\ \citenamefont {Andersson}}]{puthoor2016measurement}%
  \BibitemOpen
  \bibfield  {author} {\bibinfo {author} {\bibfnamefont {I.~V.}\ \bibnamefont
  {Puthoor}}, \bibinfo {author} {\bibfnamefont {R.}~\bibnamefont {Amiri}},
  \bibinfo {author} {\bibfnamefont {P.}~\bibnamefont {Wallden}}, \bibinfo
  {author} {\bibfnamefont {M.}~\bibnamefont {Curty}}, \ and\ \bibinfo {author}
  {\bibfnamefont {E.}~\bibnamefont {Andersson}},\ }\href@noop {} {\bibfield
  {journal} {\bibinfo  {journal} {Physical Review A}\ }\textbf {\bibinfo
  {volume} {94}},\ \bibinfo {pages} {022328} (\bibinfo {year}
  {2016})}\BibitemShut {NoStop}%
\bibitem [{\citenamefont {Hwang}(2003)}]{hwang2003quantum}%
  \BibitemOpen
  \bibfield  {author} {\bibinfo {author} {\bibfnamefont {W.-Y.}\ \bibnamefont
  {Hwang}},\ }\href@noop {} {\bibfield  {journal} {\bibinfo  {journal}
  {Physical Review Letters}\ }\textbf {\bibinfo {volume} {91}},\ \bibinfo
  {pages} {057901} (\bibinfo {year} {2003})}\BibitemShut {NoStop}%
\bibitem [{\citenamefont {Lo}\ \emph {et~al.}(2005)\citenamefont {Lo},
  \citenamefont {Ma},\ and\ \citenamefont {Chen}}]{lo2005decoy}%
  \BibitemOpen
  \bibfield  {author} {\bibinfo {author} {\bibfnamefont {H.-K.}\ \bibnamefont
  {Lo}}, \bibinfo {author} {\bibfnamefont {X.}~\bibnamefont {Ma}}, \ and\
  \bibinfo {author} {\bibfnamefont {K.}~\bibnamefont {Chen}},\ }\href@noop {}
  {\bibfield  {journal} {\bibinfo  {journal} {Physical Review Letters}\
  }\textbf {\bibinfo {volume} {94}},\ \bibinfo {pages} {230504} (\bibinfo
  {year} {2005})}\BibitemShut {NoStop}%
\bibitem [{\citenamefont {Wang}(2005)}]{wang2005beating}%
  \BibitemOpen
  \bibfield  {author} {\bibinfo {author} {\bibfnamefont {X.-B.}\ \bibnamefont
  {Wang}},\ }\href@noop {} {\bibfield  {journal} {\bibinfo  {journal} {Physical
  Review Letters}\ }\textbf {\bibinfo {volume} {94}},\ \bibinfo {pages}
  {230503} (\bibinfo {year} {2005})}\BibitemShut {NoStop}%
\bibitem [{\citenamefont {Tang}\ \emph {et~al.}(2016)\citenamefont {Tang},
  \citenamefont {Yin}, \citenamefont {Zhao}, \citenamefont {Liu}, \citenamefont
  {Sun}, \citenamefont {Huang}, \citenamefont {Zhang}, \citenamefont {Chen},
  \citenamefont {Zhang}, \citenamefont {You} \emph
  {et~al.}}]{tang2016measurement}%
  \BibitemOpen
  \bibfield  {author} {\bibinfo {author} {\bibfnamefont {Y.-L.}\ \bibnamefont
  {Tang}}, \bibinfo {author} {\bibfnamefont {H.-L.}\ \bibnamefont {Yin}},
  \bibinfo {author} {\bibfnamefont {Q.}~\bibnamefont {Zhao}}, \bibinfo {author}
  {\bibfnamefont {H.}~\bibnamefont {Liu}}, \bibinfo {author} {\bibfnamefont
  {X.-X.}\ \bibnamefont {Sun}}, \bibinfo {author} {\bibfnamefont {M.-Q.}\
  \bibnamefont {Huang}}, \bibinfo {author} {\bibfnamefont {W.-J.}\ \bibnamefont
  {Zhang}}, \bibinfo {author} {\bibfnamefont {S.-J.}\ \bibnamefont {Chen}},
  \bibinfo {author} {\bibfnamefont {L.}~\bibnamefont {Zhang}}, \bibinfo
  {author} {\bibfnamefont {L.-X.}\ \bibnamefont {You}},  \emph {et~al.},\
  }\href@noop {} {\bibfield  {journal} {\bibinfo  {journal} {Physical Review
  X}\ }\textbf {\bibinfo {volume} {6}},\ \bibinfo {pages} {011024} (\bibinfo
  {year} {2016})}\BibitemShut {NoStop}%
\bibitem [{\citenamefont {Ma}\ and\ \citenamefont
  {Razavi}(2012)}]{Ma:2012:Alternative}%
  \BibitemOpen
  \bibfield  {author} {\bibinfo {author} {\bibfnamefont {X.}~\bibnamefont
  {Ma}}\ and\ \bibinfo {author} {\bibfnamefont {M.}~\bibnamefont {Razavi}},\
  }\href@noop {} {\bibfield  {journal} {\bibinfo  {journal} {Phys. Rev. A}\
  }\textbf {\bibinfo {volume} {86}},\ \bibinfo {pages} {062319} (\bibinfo
  {year} {2012})}\BibitemShut {NoStop}%
\bibitem [{\citenamefont {Bennett}\ and\ \citenamefont
  {Brassard}(1984)}]{bennett1984quantum}%
  \BibitemOpen
  \bibfield  {author} {\bibinfo {author} {\bibfnamefont {C.~H.}\ \bibnamefont
  {Bennett}}\ and\ \bibinfo {author} {\bibfnamefont {G.}~\bibnamefont
  {Brassard}},\ }in\ \href@noop {} {\emph {\bibinfo {booktitle} {International
  Conference on Computer System and Signal Processing, IEEE}}}\ (\bibinfo
  {year} {1984})\ pp.\ \bibinfo {pages} {175--179}\BibitemShut {NoStop}%
\bibitem [{\citenamefont {Yin}\ \emph {et~al.}(2016{\natexlab{c}})\citenamefont
  {Yin}, \citenamefont {Chen}, \citenamefont {Yu}, \citenamefont {Liu},
  \citenamefont {You}, \citenamefont {Zhou}, \citenamefont {Chen},
  \citenamefont {Mao}, \citenamefont {Huang}, \citenamefont {Zhang},
  \citenamefont {Chen}, \citenamefont {Li}, \citenamefont {Nolan},
  \citenamefont {Zhou}, \citenamefont {Jiang}, \citenamefont {Wang},
  \citenamefont {Zhang}, \citenamefont {Wang},\ and\ \citenamefont
  {Pan}}]{Yin:2016:MDIQKD}%
  \BibitemOpen
  \bibfield  {author} {\bibinfo {author} {\bibfnamefont {H.-L.}\ \bibnamefont
  {Yin}}, \bibinfo {author} {\bibfnamefont {T.-Y.}\ \bibnamefont {Chen}},
  \bibinfo {author} {\bibfnamefont {Z.-W.}\ \bibnamefont {Yu}}, \bibinfo
  {author} {\bibfnamefont {H.}~\bibnamefont {Liu}}, \bibinfo {author}
  {\bibfnamefont {L.-X.}\ \bibnamefont {You}}, \bibinfo {author} {\bibfnamefont
  {Y.-H.}\ \bibnamefont {Zhou}}, \bibinfo {author} {\bibfnamefont {S.-J.}\
  \bibnamefont {Chen}}, \bibinfo {author} {\bibfnamefont {Y.}~\bibnamefont
  {Mao}}, \bibinfo {author} {\bibfnamefont {M.-Q.}\ \bibnamefont {Huang}},
  \bibinfo {author} {\bibfnamefont {W.-J.}\ \bibnamefont {Zhang}}, \bibinfo
  {author} {\bibfnamefont {H.}~\bibnamefont {Chen}}, \bibinfo {author}
  {\bibfnamefont {M.~J.}\ \bibnamefont {Li}}, \bibinfo {author} {\bibfnamefont
  {D.}~\bibnamefont {Nolan}}, \bibinfo {author} {\bibfnamefont
  {F.}~\bibnamefont {Zhou}}, \bibinfo {author} {\bibfnamefont {X.}~\bibnamefont
  {Jiang}}, \bibinfo {author} {\bibfnamefont {Z.}~\bibnamefont {Wang}},
  \bibinfo {author} {\bibfnamefont {Q.}~\bibnamefont {Zhang}}, \bibinfo
  {author} {\bibfnamefont {X.-B.}\ \bibnamefont {Wang}}, \ and\ \bibinfo
  {author} {\bibfnamefont {J.-W.}\ \bibnamefont {Pan}},\ }\href@noop {}
  {\bibfield  {journal} {\bibinfo  {journal} {Phys. Rev. Lett.}\ }\textbf
  {\bibinfo {volume} {117}},\ \bibinfo {pages} {190501} (\bibinfo {year}
  {2016}{\natexlab{c}})}\BibitemShut {NoStop}%
\bibitem [{\citenamefont {Curty}\ \emph {et~al.}(2014)\citenamefont {Curty},
  \citenamefont {Xu}, \citenamefont {Cui}, \citenamefont {Lim}, \citenamefont
  {Tamaki},\ and\ \citenamefont {Lo}}]{curty2014finite}%
  \BibitemOpen
  \bibfield  {author} {\bibinfo {author} {\bibfnamefont {M.}~\bibnamefont
  {Curty}}, \bibinfo {author} {\bibfnamefont {F.}~\bibnamefont {Xu}}, \bibinfo
  {author} {\bibfnamefont {W.}~\bibnamefont {Cui}}, \bibinfo {author}
  {\bibfnamefont {C.~C.~W.}\ \bibnamefont {Lim}}, \bibinfo {author}
  {\bibfnamefont {K.}~\bibnamefont {Tamaki}}, \ and\ \bibinfo {author}
  {\bibfnamefont {H.-K.}\ \bibnamefont {Lo}},\ }\href@noop {} {\bibfield
  {journal} {\bibinfo  {journal} {Nature communications}\ }\textbf {\bibinfo
  {volume} {5}} (\bibinfo {year} {2014})}\BibitemShut {NoStop}%
\bibitem [{\citenamefont {Brassard}\ and\ \citenamefont
  {Salvail}(1993)}]{brassard1993secret}%
  \BibitemOpen
  \bibfield  {author} {\bibinfo {author} {\bibfnamefont {G.}~\bibnamefont
  {Brassard}}\ and\ \bibinfo {author} {\bibfnamefont {L.}~\bibnamefont
  {Salvail}},\ }in\ \href@noop {} {\emph {\bibinfo {booktitle} {Workshop on the
  Theory and Application of of Cryptographic Techniques}}}\ (\bibinfo
  {organization} {Springer},\ \bibinfo {year} {1993})\ pp.\ \bibinfo {pages}
  {410--423}\BibitemShut {NoStop}%
\bibitem [{\citenamefont {Serfling}(1974)}]{serfling1974probability}%
  \BibitemOpen
  \bibfield  {author} {\bibinfo {author} {\bibfnamefont {R.~J.}\ \bibnamefont
  {Serfling}},\ }\href@noop {} {\bibfield  {journal} {\bibinfo  {journal} {The
  Annals of Statistics}\ ,\ \bibinfo {pages} {39}} (\bibinfo {year}
  {1974})}\BibitemShut {NoStop}%
\bibitem [{\citenamefont {Comandar}\ \emph {et~al.}(2016)\citenamefont
  {Comandar}, \citenamefont {Lucamarini}, \citenamefont {Fr{\"o}hlich},
  \citenamefont {Dynes}, \citenamefont {Sharpe}, \citenamefont {Tam},
  \citenamefont {Yuan}, \citenamefont {Penty},\ and\ \citenamefont
  {Shields}}]{comandar2016quantum}%
  \BibitemOpen
  \bibfield  {author} {\bibinfo {author} {\bibfnamefont {L.}~\bibnamefont
  {Comandar}}, \bibinfo {author} {\bibfnamefont {M.}~\bibnamefont
  {Lucamarini}}, \bibinfo {author} {\bibfnamefont {B.}~\bibnamefont
  {Fr{\"o}hlich}}, \bibinfo {author} {\bibfnamefont {J.}~\bibnamefont {Dynes}},
  \bibinfo {author} {\bibfnamefont {A.}~\bibnamefont {Sharpe}}, \bibinfo
  {author} {\bibfnamefont {S.-B.}\ \bibnamefont {Tam}}, \bibinfo {author}
  {\bibfnamefont {Z.}~\bibnamefont {Yuan}}, \bibinfo {author} {\bibfnamefont
  {R.}~\bibnamefont {Penty}}, \ and\ \bibinfo {author} {\bibfnamefont
  {A.}~\bibnamefont {Shields}},\ }\href@noop {} {\bibfield  {journal} {\bibinfo
   {journal} {Nature Photonics}\ } (\bibinfo {year} {2016})}\BibitemShut
  {NoStop}%
\end{thebibliography}


%

\end{document}